\documentclass[11pt,a4paper]{article}

\usepackage[T1]{fontenc}
\usepackage[utf8]{inputenc}
\usepackage[UKenglish]{babel}

\usepackage[margin=2.5cm]{geometry}

\usepackage{mathtools,amssymb}
\usepackage{graphicx,tikz,physics,dsfont}
\usepackage{booktabs,array,subcaption}
\usepackage{cite}
\usepackage[justification=raggedright]{caption}
\usepackage{hyperref}
\usepackage{eso-pic}

\newcommand{\oo}{\mathrm{O}}

\title{
Subtraction method for disconnected diagrams in full QCD
}

\author{
Lorenzo Barca$^{1}$,
Joshua Crawford$^{2}$,
Jacob Finkenrath$^{2}$,
Stefan Schaefer$^{1}$,\\
Juan Andr\'es Urrea-Ni\~no$^{2,3,4}$
\\[1em]
\small $^1$John von Neumann-Institut für Computing NIC,
Deutsches Elektronen-Synchrotron DESY,\\
\small Platanenallee 6, 15738 Zeuthen, Germany
\\[0.5em]
\small $^2$Department of Physics,
Bergische Universität Wuppertal,\\
\small Gaußstr. 20, 42119 Wuppertal, Germany
\\[0.5em]
\small $^3$School of Mathematics,
Trinity College Dublin, Dublin 2, Ireland
\\[0.5em]
\small $^4$Hamilton Mathematics Institute,
Trinity College Dublin,\\
\small College Green, Dublin 2, Ireland
}

\date{}

\begin{document}
\AddToShipoutPictureFG*{%
  \AtPageUpperLeft{%
    \put(\LenToUnit{\dimexpr\paperwidth-2.5cm\relax},
         \LenToUnit{-2.0cm}){%
      \makebox[0pt][r]{%
        \begin{tabular}{r}
          \texttt{DESY-26-127}\\
          \texttt{WUB/26-04}
        \end{tabular}
      }%
    }%
  }%
}

\maketitle

\begin{abstract}
Disconnected diagrams remain a major challenge for precision lattice QCD
calculations because of their large statistical fluctuations.
We investigate a subtraction method based on a domain decomposition of
the Dirac operator, yielding a biased but substantially less noisy
estimator. Applying the method to charmonium disconnected correlators
computed using distillation, we achieve a variance reduction of about
three orders of magnitude at only twice the computational cost,
providing a practical alternative to completely neglecting
disconnected contributions.
\end{abstract}

\section{Introduction}
Disconnected diagrams play an important role in many lattice QCD observables of flavour-singlet hadrons. Their numerical evaluation, however, remains one of the most challenging problems in lattice QCD due to the extremely large statistical fluctuations associated with quark loops. In disconnected correlators, the variance is typically dominated by vacuum expectation values of quark-loop operators \cite{Parisi:1983ae,Lepage:1989hd}, leading to a signal-to-noise ratio that deteriorates rapidly with Euclidean time separation.

The problem becomes particularly severe in systems involving heavy quarks as the signal of a hadronic correlator decreases exponentially faster than with light quarks. In charmonium spectroscopy, charm quark-line disconnected diagrams, associated with charm annihilation, are often neglected because they are expected to be suppressed by the Okubo--Zweig--Iizuka (OZI) rule. Their quantitative impact, however, remains poorly understood outside the perturbative regime and represents an important source of uncertainty in precision determinations of the charmonium spectrum, as illustrated by the still unsettled disconnected contribution to the $\eta_c$ mass \cite{Urrea-Nino:2025afu}.
A similar limitation arises in disconnected diagrams of nucleon three-point functions, where the exponentially worsening signal-to-noise ratio restricts source-sink separations and can leave sizeable excited-state contamination \cite{Barca:2025det,Barca:2024hrl}.

A number of techniques have been developed to mitigate the large fluctuations of disconnected diagrams. These include improved stochastic estimators \cite{Bali:2009hu}, frequency splitting \cite{Giusti:2019kff}, dilution schemes \cite{Bali:2021qem}, low-mode averaging \cite{Gruber:2024cos,Neff:2001zr,DeGrand:2004qw}, recent developments of noise-reduction techniques with machine learning \cite{Abbott:2026vui,Abbott:2026ylv} and multi-level algorithms exploiting locality in lattice QCD \cite{Barca:2025dca,Barca:2024fpc,Ce:2016idq,Ce:2016ajy}. Despite these advances, the computational cost of reducing the statistical errors for a good signal-to-noise remains still high.

In this work we propose a simple method to reduce the
fluctuations of disconnected two-point correlators by
subtracting a term that is strongly correlated with the full
disconnected correlator. This term is constructed from quark
loops restricted to non-overlapping regions and captures a
large fraction of the statistical fluctuations. Its contribution
to the physical signal is nevertheless expected to be small
in some mesonic channels, since the two fermionic loops
communicate only through gauge-field fluctuations in the sea.
The subtraction therefore removes the dominant noise contribution while retaining
the terms in which the quark propagators overlap through common regions of the lattice.
The resulting estimator is biased,
with the size of the bias determined by the contribution of
the subtracted term to the physical signal. This provides a
practical alternative to completely neglecting disconnected
diagrams. We investigate how the error reduction and the
induced bias depend on the size of the non-overlapping
regions and on the channel considered.
While this work was in preparation, a similar approach exploiting the locality of quark propagation was independently applied to the study of the $\eta$ and $\eta'$ mesons and presented by Luchang Jin at the Lattice 2026 conference~\cite{Jin:2026}.

We test the method on flavour-singlet charmonium two-point functions and find a
substantial reduction of the statistical fluctuations at modest additional
computational cost. 

\section{Methodology}
Quark-line disconnected diagrams arise in correlation functions of flavour-singlet operators
\begin{equation}
\oo_\Gamma(x) = \bar q(x) \Gamma q(x),
\end{equation}
with a single quark flavor $q$, where $x=(x_0,\vec{x})$ and whose two-point correlation function can be decomposed into the quark-line connected $C_\Gamma^{\rm (c)}$ and disconnected $C_\Gamma^{\rm (d)}$ contributions, 
\begin{equation}
\label{Cfull}
C_\Gamma(x_0,y_0) = C_\Gamma^{\rm (c)}(x_0,y_0) + C_\Gamma^{\rm (d)}(x_0,y_0).
\end{equation}
Explicitly, the disconnected contributions read
\begin{align}
\label{Cdisc}
C^{\rm (d)}_\Gamma(x_0,y_0) = &~\langle L_\Gamma(x_0) {L}_{\widetilde{\Gamma}}(y_0)\rangle
- \langle L_\Gamma(x_0)\rangle~ \langle L_{\widetilde{\Gamma}}(y_0)\rangle
\end{align}
where, for simplicity, we define the quark loop as $L_\Gamma (x_0) = \sum_{\vec{x}} \mathrm{Tr} \left[\Gamma D^{-1}(x,x)\right]$, $\widetilde{\Gamma}=\gamma_0 \Gamma^\dagger \gamma_0$, 
consider operators projected to zero momentum, and denote the gauge average by $\langle \cdot \rangle$.
In Eq.~\eqref{Cdisc}, we subtract the vacuum expectaction value (VEV) to consider only the gauge-connected piece.
In numerical simulations, the variance of the correlator in Eq.~\eqref{Cfull} is typically dominated at sufficiently long distances by fluctuations of the VEV in the four-point functions, i.e.,
\begin{align}
\label{variance}
\sigma^2_{C}(x_0, y_0)
=&
~\sum_{\vec{x}, \vec{y}}\langle \left(\oo(x) \oo(y)\right)^2\rangle 
-
C(x_0, y_0)^2
\\
~\approx &~
\langle L_\Gamma(x_0)^2 \rangle ~\langle L_{\widetilde{\Gamma}}(y_0)^2\rangle
= ~\mathrm{const.}
\end{align}
In particular, in a free field theory, one can show that the VEV is proportional to the spatial volume $L^3$~\cite{Lepage:1989hd}, due to the momentum projection. As a consequence, in standard simulations, the signal-to-noise (S/N) deteriorates exponentially as $\mathrm{exp}(-m_\Gamma |x_0-y_0|)\sqrt{N}$ with $m_\Gamma$ the mass of the lightest state with the quantum numbers corresponding to the operator $\oo_\Gamma$, and $N$ the number of Monte Carlo samples.

\subsection{Domain decomposition of the propagator}
\begin{figure}[t]
\centering
\includegraphics[width=0.7\linewidth]{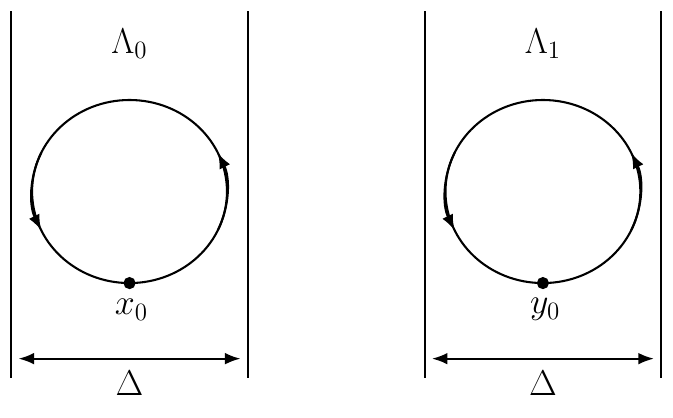}
\caption{Representation of the first term in Eq.~\eqref{correlator_factorised}, where $\Delta$ is the size of the non-overlapping temporal regions $\Lambda_0$ and $\Lambda_1$.}
\label{fig:block_correlator}
\end{figure}

To exploit the locality of quark propagation, we partition the temporal lattice extent $\Lambda$ into two non-overlapping regions $\Lambda_i$ and $\Lambda_i^*$, i.e.,
\begin{equation}
    \Lambda = \Lambda_i \cup \Lambda_i^*.
\end{equation}
Following Ref.~\cite{Ce:2016idq}, the Dirac operator can be written in block form,
\begin{equation}
D =
\begin{pmatrix}
D_{\Lambda_i} & D_{\partial \Lambda_{i}} \\
D_{\partial \Lambda_{i}^*} & D_{\Lambda^*_i}
\end{pmatrix}.
\end{equation}
The operator $D_{\Lambda_i}$($D_{\Lambda_i^*}$) is obtained by restricting $D$ to the quark fields inside $\Lambda_i$ ($\Lambda_i^*$), by imposing Dirichlet boundary conditions on $\partial \Lambda_i$ ($\partial \Lambda^*_i$).
Using this domain decomposition, the exact quark propagator $D^{-1}(x, x)$ with $x=(x_0, \vec{x})$ and $x_0\in \Lambda_i$, can be written as
\begin{equation}
\label{factorisation}
D^{-1}(x,x)
=
D^{-1}_{\Lambda_i}(x,x)
+
D^{-1}_{\Lambda_i}
\left[ D_{\partial \Lambda_i} D^{-1}D_{\partial \Lambda_i^*}\right]
D^{-1}_{\Lambda_i}
.
\end{equation}
The first term describes quark propagation entirely within $\Lambda_i$, 
see Fig.~\ref{fig:block_correlator} for the quark loops in the regions $\Lambda_0$ and $\Lambda_1$,
while the remaining terms correspond to propagation that leaves the region $\Lambda_i$
and re-enters through the boundaries.

\subsection{Factorized contributions}
Using this decomposition, the quark loop is separated into a contribution 
local to the region and corrections involving propagation through other regions,
\begin{equation}
\label{quark_loop_factorised}
L_\Gamma(x_0) = L_{\Gamma, \Lambda_0}(x_0) + \delta L_{\Gamma, \Lambda_0}(x_0),
\end{equation}
where $L_{\Lambda_0}$ and $\delta L_{\Lambda_0}$ are the zero-momentum traces of the two terms in the factorisation in Eq.~\eqref{factorisation}.
As the size of $\Lambda_0$ is increased, $L_{\Gamma,\Lambda_0}$ provides an
increasingly accurate approximation to the full quark loop $L_\Gamma$, while
the correction $\delta L_{\Gamma,\Lambda_0}$ becomes increasingly suppressed. 
This behaviour was verified numerically in Ref.~\cite{Barca:2025dca}, where the 
corrections to the VEV were found to decrease exponentially with the distance
between the operator and the boundaries of the region, i.e., with the size of the regions.
Using this factorisation, the disconnected term in Eq.~\eqref{Cdisc} reads
\begin{equation}
\label{correlator_factorised}
C^{\rm (d)}_\Gamma(x_0, y_0)
=
C^{\rm(d)}_{\Gamma, \Lambda_0, \Lambda_1}(x_0, y_0)
+
\delta C^{\rm(d)}_{\Gamma, \Lambda_0, \Lambda_1}(x_0, y_0),
\end{equation}
where the first term is the VEV-subtracted, completely factorised disconnected contribution, i.e.,
\begin{equation}
C^{\rm(d)}_{\Gamma, \Lambda_0, \Lambda_1}(x_0, y_0) = 
\langle L_{\Gamma, \Lambda_0}(x_0) L_{\widetilde{\Gamma}, \Lambda_1}(y_0) \rangle
-
\langle L_{\Gamma, \Lambda_0}(x_0)\rangle \langle L_{\widetilde{\Gamma}, \Lambda_1}(y_0) \rangle.
\end{equation}
The main intuition is that, when $\Lambda_0$ and $\Lambda_1$ are non-overlapping 
and $x_0$ and $y_0$ are sufficiently far from the boundaries,
$C^{\rm(d)}_{\Gamma,\Lambda_0,\Lambda_1}$ is expected to be small for fermionic states
compared to $\delta C^{\rm(d)}_{\Gamma,\Lambda_0,\Lambda_1}$. The corresponding fermionic
lines do not propagate through a common region, so their correlation is
mediated only through gauge-field fluctuations in the sea.
At fixed source-sink separation $|x_0-y_0|$, decreasing the size of the
regions increases their separation $d(\Lambda_0,\Lambda_1)$ and therefore
further suppresses this contribution. At sufficiently large separations, we
expect $C^{\rm(d)}_{\Gamma,\Lambda_0,\Lambda_1} \sim
e^{-m_\Gamma^{\rm sea}\,d(\Lambda_0,\Lambda_1)}$,
where $m_\Gamma^{\rm sea}$ denotes the lowest energy scale with the appropriate
quantum numbers that can mediate the correlation between the two regions.
Fig.~\ref{fig:block_correlator} shows the representation of the quark loops $L_{\Gamma, \Lambda_0}(x_0)$ and $L_{\Gamma, \Lambda_1}(y_0)$ in the regions $\Lambda_0$ and $\Lambda_1$, respectively.

Regarding the statistical fluctuations, using the decomposition in Eq.~\eqref{quark_loop_factorised}, the constant and dominant term in the variance in Eq.~\eqref{variance} reads
\begin{align}
\sigma^2_{C}(x_0, y_0)
&\approx
\nonumber
\langle L_{\Gamma, \Lambda_0}(x_0)^2\rangle ~\langle L_{\widetilde{\Gamma}, \Lambda_1}(y_0)^2\rangle
\\
\nonumber
&+
\langle L_{\Gamma, \Lambda_0}(x_0)^2\rangle ~\langle \delta L_{\widetilde{\Gamma}, \Lambda_1}(y_0)^2\rangle
\\
\nonumber
&+
\label{variance_factorised}
\langle \delta L_{\Gamma, \Lambda_0}(x_0)^2\rangle ~\langle L_{\widetilde{\Gamma}, \Lambda_1}(y_0)^2\rangle
\\
&+
\langle \delta L_{\Gamma, \Lambda_0}(x_0)^2\rangle ~ \langle \delta L_{\widetilde{\Gamma}, \Lambda_1}(y_0)^2\rangle.
\end{align}
As discussed above, as the sizes of $\Lambda_0$ and $\Lambda_1$ are increased,
$L_{\Gamma,\Lambda_i}$ provides an increasingly accurate approximation to the
full quark loop. Consequently, the first term in
Eq.~\eqref{variance_factorised} approaches the full VEV contribution to the
variance, while the remaining terms, which contain at least one
$\delta L_{\Gamma,\Lambda_i}$, become increasingly suppressed.

\subsection{Subtraction method}
The observations above motivate subtracting the completely factorised
contribution from the full disconnected correlator in Eq.~\eqref{correlator_factorised}. 
We therefore consider
\begin{equation}
\label{subtracted_correlator}
\delta C^{\rm(d)}_{\Gamma,\Lambda_0,\Lambda_1}(x_0,y_0)
=
C^{\rm(d)}_\Gamma(x_0,y_0)
-
C^{\rm(d)}_{\Gamma,\Lambda_0,\Lambda_1}(x_0,y_0),
\end{equation}
whose variance is free of the first term in Eq.~\eqref{variance_factorised}. 
The resulting estimator is biased by the neglected completely factorised contribution.

The size of the regions controls a trade-off between statistical precision
and bias. Increasing the sizes of $\Lambda_0$ and $\Lambda_1$ makes
$L_{\Gamma,\Lambda_0}(x_0)$ and $L_{\Gamma,\Lambda_1}(y_0)$ 
better approximations to the full quark loops $L_\Gamma(x_0)$ and
$L_\Gamma(y_0)$, respectively, see Fig.~2 in Ref.~\cite{Barca:2025dca}. 
Consequently, the completely factorised term
captures an increasingly large fraction of the dominant fluctuations, and its
subtraction leads to an exponential reduction of the variance.
At fixed source-sink separation, however, larger regions
reduce the distance $d(\Lambda_0,\Lambda_1)$ between them, thereby increasing
the neglected completely factorised contribution and hence the bias.
An optimal choice of the region size must therefore balance these two effects,
unless the neglected contribution is known to be irrelevant for the
observable of interest.

Importantly, the bias associated with the completely factorised contribution
$C^{\rm(d)}_{\Gamma,\Lambda_0,\Lambda_1}(x_0,y_0)$ can be determined more
efficiently using two-level sampling techniques, as demonstrated in
Refs.~\cite{Barca:2025dca,Barca:2024fpc,Ce:2016idq,Ce:2016ajy}.

\section{Results}
We consider an ensemble with $N_f=3+1$ flavours in the sea and the Wilson-Clover action with open boundary conditions in time \cite{Luscher:2011kk} and study the subtraction method using the charm isosinglet interpolating operators $\mathrm{O}^{c\bar{c}}_\Gamma = \bar{c} \Gamma c$.
The lattice extent is $V/a^4=96\times 32^3$ with $a=0.054~\rm fm$ and $m_\pi = 420~\rm MeV$, and we analyse $2000$ gauge configurations; see Refs.~\cite{Urrea-Nino:2025afu,Hollwieser:2020qri} for more details.

\subsection{Analysis with standard distillation}
We construct the quark sources using distillation \cite{HadronSpectrum:2009krc}, with $200$ Laplacian eigenvectors computed on APE-smeared gauge links, see Ref.~\cite{Urrea-Nino:2025afu} for the smearing parameters.

For the computation of the completely factorised disconnected term $C^{(d)}_{\Gamma, \Lambda_0, \Lambda_1}(x_0, y_0)$,
we consider $\Lambda_0$ and $\Lambda_1$ to be of the same width $\Delta$.
The argument that the signal of $C^{\rm(d)}_{\Gamma, \Lambda_0, \Lambda_1}$ is small is valid only when $|x_0-y_0| \geq \Delta$, i.e., when the regions $\Lambda_0, \Lambda_1$ are non-overlapping and the fermionic lines do not come close. In this scenario, the contribution is arising from the sea.

\begin{figure}[!htbp]
\centering
\includegraphics[width=0.9\linewidth]{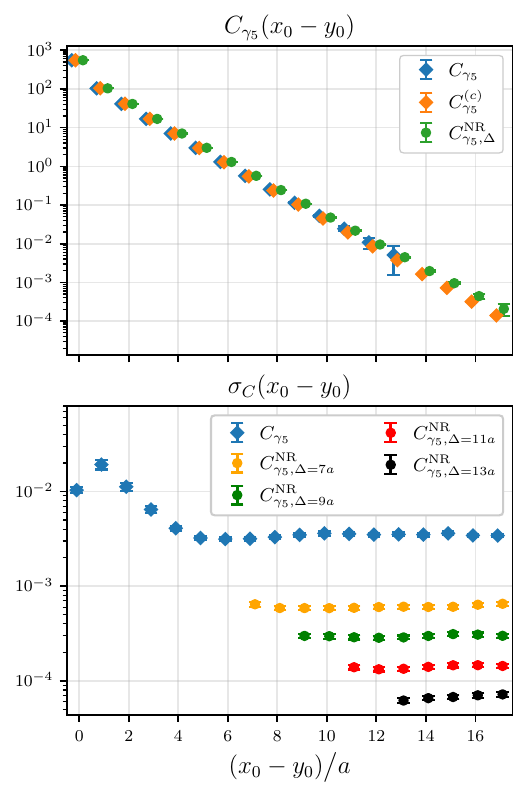}
\caption{(top) Comparison between the full two-point correlation function in Eq.~\eqref{Cfull} with $\Gamma=\gamma_5$ and the noise-reduced correlator $C_{\gamma_5, \Delta}^{\rm NR}$. The connected contribution is included for reference. (bottom) Statistical errors of the full and noise-reduced estimators with different block sizes.}
\label{fig:corr_blocks_G5}
\end{figure}

In Fig.~\ref{fig:corr_blocks_G5}, we show the results for the full correlation function (top plot)
and its statistical errors (bottom plot) of $\mathrm{O}^{c\bar{c}}_{\gamma_5}(x_0)$ ($\eta_c$ channel),
using 4 block sizes $\Delta/a = 7,~9,~11$, and $13$ for $\Lambda_0$ and $\Lambda_1$.
The results are averaged over different source positions $y_0$. For convenience, 
we define the corresponding noise-reduced estimator for Eq.~\eqref{Cfull} as
\begin{equation}
C^{\rm NR}_{\Gamma, \Delta}(x_0,y_0)
\equiv 
C^{\rm(c)}_\Gamma(x_0,y_0) + \delta C^{\rm(d)}_{\Gamma, \Lambda_0, \Lambda_1}(x_0,y_0).
\end{equation}
For the final noise-reduced estimator $C^{\rm NR}_{\Gamma, \Delta}$ in the top plot in Fig.~\ref{fig:corr_blocks_G5}, we consider the block $\Delta=7a$ for distances $x_0-y_0=8a,9a$,
$\Delta=9a$ for $x_0-y_0=10a,11a$, $\Delta=11a$ for $x_0-y_0= 12a,13a$, and $\Delta=13a$ for $x_0-y_0\geq 14a$.

We find that the statistical errors of the disconnected contribution, and
consequently of the full correlator, are reduced by factors that increase with
the block size. For instance, for $\Delta/a=13$, the statistical error is
reduced by almost a factor of $50$ compared to the standard disconnected
estimator $C^{(d)}_\Gamma$. This extends the range over which a statistically
significant signal can be resolved in both the full correlator, cf. the top
panel of Fig.~\ref{fig:corr_blocks_G5}, and the corresponding effective mass,
see Fig.~\ref{fig:meff_blocks}, by about five lattice sites.

The statistical errors of the subtracted correlator in Eq.~\eqref{subtracted_correlator} decrease by a factor
of $\approx2$ when we increase the block width by $\delta \Delta/a=2$. 
This error reduction can be explained by noticing that the fluctuations of the subtracted correlator are dominated by the second and third terms in Eq.~\eqref{variance_factorised}.
As we increase the width of $\Lambda_0$ and $\Lambda_1$ by $\delta \Delta/a=2$ each, we find 
that the statistical errors reduce by a factor of $\approx 2$, see bottom plot in Fig.~\ref{fig:corr_blocks_G5}.

\begin{figure}[!htbp]
\centering
\includegraphics[width=0.9\linewidth]{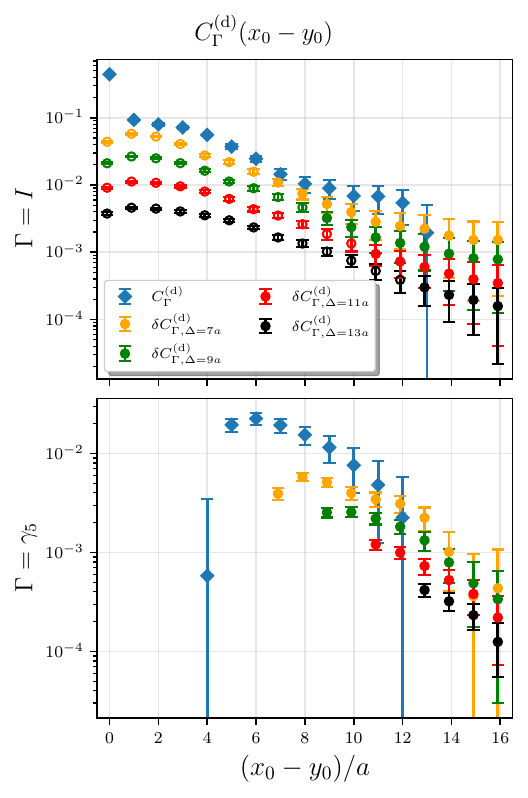}
\caption{Comparison of the disconnected contributions of the two-point correlation functions between 
the unbiased correlator and the subtracted correlator in Eq.~\eqref{subtracted_correlator} with different block sizes $\Delta=7a,~9a,~11a$, and $13a$, and for $\Gamma=\gamma_5$, and $\Gamma=I$.
For the latter, we also include for reference the data points of the noise-reduced disconnected term for $x_0-y_0< \Delta$ (empty circles). 
}
\label{fig:corr_blocks}
\end{figure}

\begin{figure}[!htbp]
\centering
\includegraphics[width=0.9\linewidth]{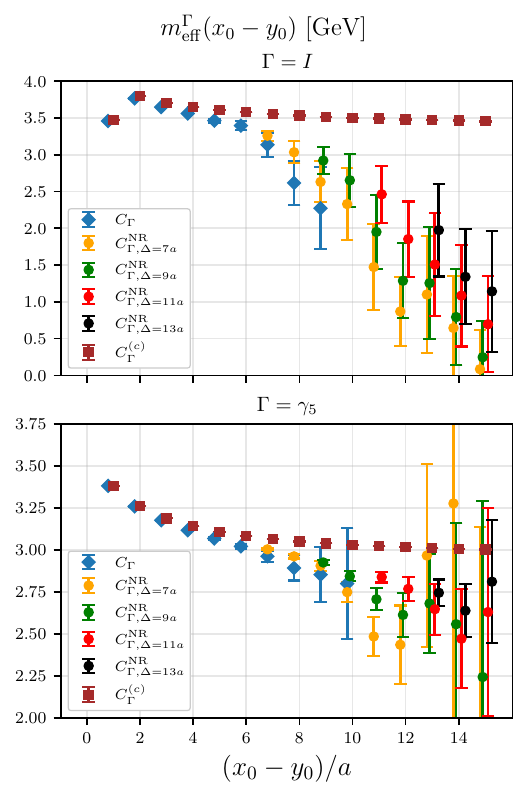}
\caption{Comparison of the effective masses for the full two-point correlation functions between 
the unbiased estimator $C_\Gamma$ and the biased noise-reduced estimators $C_{\Gamma, \Delta}^{\rm NR}$ with $\Delta/a=7,~9,~11$ and $13$, and for $\Gamma=I,\gamma_5$.}
\label{fig:meff_blocks}
\end{figure}

However, this subtraction introduces a bias. The neglected term $C^{\rm(d)}_{\Gamma, \Lambda_0, \Lambda_1}(x_0, y_0)$ for the subtracted correlator in Eq.~\eqref{subtracted_correlator} depends on the channel, interpolating operators, and on the tunable parameter $\Delta$, i.e., the block size.
In Fig.~\ref{fig:corr_blocks}, there is a comparison between the full disconnected correlator $C_\Gamma^{\rm (d)}$ and the subtracted correlator $\delta C_{\Gamma, \Delta}^{\rm (d)}$ for $\Gamma=I$, $\gamma_5$ and with $\Delta/a=7$, $9$, and $11$.
As the source-sink separation increases, the difference between the two becomes exponentially smaller.
As expected, the difference is large for distances $x_0-y_0<\Delta$, because in this case, the blocks overlap and thus the quark loops are in the same block.
In order to retrieve the full correlator with good statistical errors, the bias $C_{\Gamma, \Lambda_0, \Lambda_1}$ can be estimated via a two-level sampling technique, as demonstrated in Refs.~\cite{Barca:2025dca,Ce:2016ajy}. With this better sampling method, the dominant statistical errors contributing to $C_{\Gamma, \Lambda_0, \Lambda_1}$ decrease with the number of Monte Carlo submeasurements.

In Fig.~\ref{fig:meff_blocks}, we show the effective masses of the unbiased two-point function $C_\Gamma$ and the biased noise-reduced estimators $C_{\Gamma,\Delta}^{\rm NR}$ for $\Gamma=\gamma_5,\,I$ and different values
of $\Delta$. As a reference, we also include the effective mass of the connected contribution. The reduced statistical uncertainties allow us to resolve the correlators at larger Euclidean time separations, where a noticeable difference between the noise-reduced and connected effective masses emerges. In the $\eta_c$ channel, such a mass shift can play an important role in the hyperfine splitting \cite{Urrea-Nino:2025afu,Brambilla:2019esw}.

However, the accessible Euclidean time separations are not yet sufficiently large to establish ground-state dominance and identify a single-state plateau. This is particularly evident in the $\Gamma=I$ channel, see Fig.~\ref{fig:meff_blocks}, where the effective masses continue to decrease with increasing time separation, indicating that contributions from lower-lying states are significant. A dedicated variational analysis employing a basis of operators with overlap onto the relevant states is therefore required to disentangle their contributions and reliably determine the energy spectrum.

\subsection{Analysis with Gaussian-profiled distillation}
\begin{figure}[!htbp]
\centering
\includegraphics[width=0.9\linewidth]{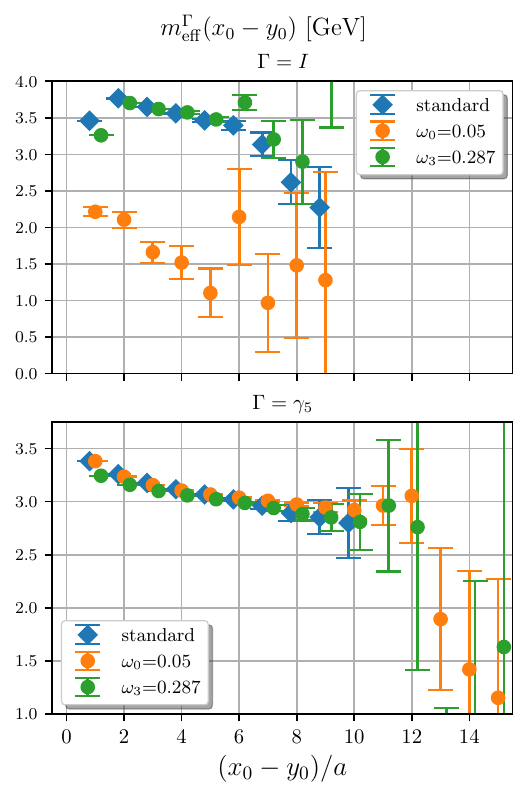}
\caption{Comparison of the effective masses for the two-point correlation functions with different Gaussian profile widths. The standard profile corresponds to the flat profile normally used in standard distillation.}
\label{fig:meff_profiles}
\end{figure}

In addition to standard distillation, following Refs.~\cite{Urrea-Nino:2026cjj,Knechtli:2022bji}, 
we construct profile-based sources for the charmonium operators.
The distillation eigenvectors $v_i(x_0)$ are eigenvectors of the gauge-covariant
three-dimensional Laplacian on the time slice $x_0$, with corresponding
eigenvalues $\lambda_i(x_0)$. Rather than weighting all retained eigenmodes
equally, as in standard distillation, we modulate their contribution using
four Gaussian profiles~\cite{Knechtli:2022bji},
\begin{equation}
g_k(\lambda_i)=\exp\left(-\frac{\lambda_i^2}{2(0.8\omega_k)^2}\right),
\end{equation}
with widths $\omega_k=0.05,~0.129,~0.208$, and $0.287$.
For interpolating operators of the form $\bar c\Gamma c$, where $\Gamma$
acts only in Dirac space and no derivatives are present, orthogonality of
the Laplacian eigenvectors simplifies the elementals to
\begin{equation}
\phi^{ij}_{\alpha\beta,k}(x_0)
=
\Gamma_{\alpha\beta}
|g_k(\lambda_i(x_0))|^2\,\delta_{ij},
\end{equation}
where $i,j$ denote the Laplacian-eigenvector indices,
$\alpha,\beta$ the Dirac indices, and $k$ labels the four Gaussian profiles.
Smaller Gaussian widths suppress the higher Laplacian modes more strongly, effectively reducing the number of eigenvectors contributing to the interpolating operator and resulting in a more spatially extended smearing profile. Conversely, larger widths retain a broader range of eigenmodes, yielding a more spatially localized operator.
In Fig.~\ref{fig:meff_profiles}, we compare the effective masses
\begin{equation}
am_{\rm eff}^\Gamma(x_0-y_0) = \ln\frac{C_{\Gamma}(x_0-y_0)}{C_{\Gamma}(x_0-y_0+a)}
\end{equation}
obtained with standard and Gaussian-profiled distillation, using the smallest and largest Gaussian widths, $\omega_0=0.05$ and $\omega_3=0.287$, respectively.

Although we analyse all four Gaussian profiles, we show only the two
extreme widths, $\omega_0=0.05$ and $\omega_3=0.287$, to illustrate
the dependence on the profile width, as the intermediate profiles
yield qualitatively similar results.

In the $\Gamma=I$ channel, the smallest Gaussian width enhances the overlap with lower-lying states, leading to an effective mass around $1.5~\rm GeV$, while in the $\Gamma=\gamma_5$ channel, it reduces the statistical noise. Such contributions from lighter states arise because the charm flavour-singlet operator can couple, through disconnected diagrams, to lighter flavour-singlet $q\bar q$ states, glueballs, and multi-hadron states. This behaviour has also been observed in Ref.~\cite{UrreaNiño:20251g} using a dedicated GEVP analysis.
We then apply the subtraction method in Eq.~\eqref{subtracted_correlator} to the quark-line disconnected correlator with the four Gaussian profiles.

\begin{figure}[!htbp]
\centering
\includegraphics[width=0.9\linewidth]{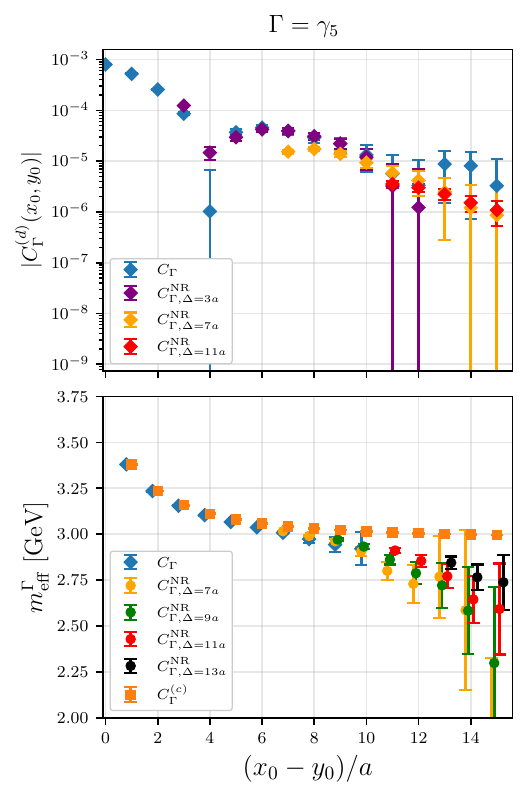}
\caption{Disconnected correlators (top) and effective masses of the full correlators (bottom), obtained by combining Gaussian-profiled distillation with the noise-subtraction method introduced in this work. The correlators are computed using the narrowest Gaussian profile, with width $\omega_0=0.05$.}
\label{fig:corr_meff_g5_block_profiles}
\end{figure}

In the top panel of Fig.~\ref{fig:corr_meff_g5_block_profiles}, we show the disconnected contribution
for the narrowest profile, $\omega_0=0.05$, and different block sizes $\Delta$.
We include also $\Delta/a=3$ to illustrate that, at comparable source-sink separations, 
the bias is substantially reduced compared to larger block sizes.
In the bottom panel, we show the corresponding effective masses. A comparison with 
the bottom plot in Fig.~\ref{fig:meff_blocks} shows that, at fixed block size, 
the narrowest Gaussian profile provides an additional reduction of approximately a factor 
of two in the statistical uncertainty of the effective masses.

In the $\Gamma=I$ channel, instead, we find that most of the signal is contained in the
completely factorised block-block correlator $C^{\rm(d)}_{\Gamma, \Lambda_0, \Lambda_1}$, 
especially for the narrowest Gaussian profile, $\omega_0=0.05$. Neglecting this contribution therefore removes 
most of the physical signal and renders the subtraction method ineffective in this case.

\begin{figure}[!htbp]
\centering
\includegraphics[width=0.9\linewidth]{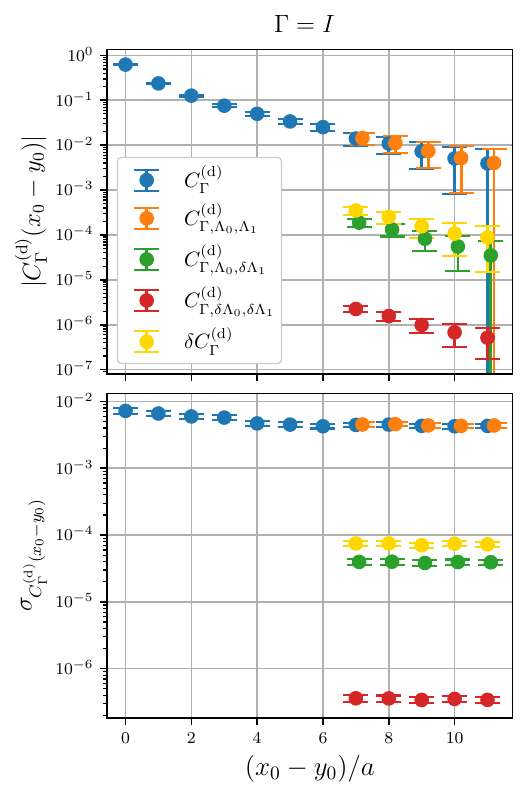}
\caption{Plots of the disconnected correlators (top) and their statistical errors (bottom) for the combinations in Eqs.~\eqref{corr1}-\eqref{corr4}. The displayed two-point functions are computed with the Gaussian profile of width $\omega_0=0.05$.}
\label{fig:corr_error_block_profiles}
\end{figure}

To show the different contributions, we compute the VEV-subtracted terms that decompose $C^{\rm (d)}_\Gamma$, namely
\begin{align}
\label{corr1}
C^{\rm(d)}_{\Lambda_0, \Lambda_1} 
&= 
\langle L_{\Lambda_0}(x_0) L_{\Lambda_1}(y_0) \rangle
-
\langle L_{\Lambda_0}(x_0)\rangle \langle L_{\Lambda_1}(y_0)\rangle;
\\
C^{\rm(d)}_{\Lambda_0, \delta \Lambda_1} 
&= 
\langle L_{\Lambda_0}(x_0) \delta L_{\Lambda_1}(y_0) \rangle
-
\langle L_{\Lambda_0}(x_0)\rangle \langle \delta L_{\Lambda_1}(y_0)\rangle;
\end{align}
\begin{align}
C^{\rm(d)}_{\delta \Lambda_0, \Lambda_1} 
&= 
\langle \delta L_{\Lambda_0}(x_0) L_{\Lambda_1}(y_0) \rangle
-
\langle \delta L_{\Lambda_0}(x_0)\rangle \langle L_{\Lambda_1}(y_0)\rangle;
\\
\label{corr4}
C^{\rm(d)}_{\delta \Lambda_0, \delta \Lambda_1} 
&= 
\langle \delta L_{\Lambda_0}(x_0) \delta L_{\Lambda_1}(y_0) \rangle
-
\langle \delta L_{\Lambda_0}(x_0)\rangle \langle \delta L_{\Lambda_1}(y_0)\rangle,
\end{align}
where we remove the subscript $\Gamma$ to shorten the notation.
In Fig.~\ref{fig:corr_error_block_profiles}, we show the two-point correlation functions in Eqs.~\eqref{corr1}-\eqref{corr4} for $\Gamma=I$, along with their statistical errors. For this comparison, we consider the block size $\Delta=7a$. 
We find that the block-block correlator $C^{\rm(d)}_{\Lambda_0,\Lambda_1}$ contains most of both the signal and the statistical noise, as shown in the bottom panel. Consequently, neglecting this contribution introduces a significantly larger bias in the subtracted correlator $\delta C_I^{\rm(d)}$ than in the case of standard distillation shown in Fig.~\ref{fig:corr_blocks}.

However, the fact that both the signal and the statistical noise are concentrated in the block-block correlator makes this contribution particularly suitable for multilevel sampling. By computing this dominant contribution with a multilevel algorithm and the remaining, less noisy terms with standard sampling, one can reduce the statistical uncertainties while retaining an unbiased estimator of the full correlator. This approach also avoids the need for a more elaborate approximation of the quark propagator involving additional terms, providing a promising direction for further improving the efficiency of the method \cite{Barca:2025dca}.

\section{Cost}
The additional cost associated with this method is the computation of the factorised quark loops
$L_{\Gamma, \Lambda_0}(x_0)$, and $L_{\Gamma, \Lambda_1}(y_0)$, because the correction term in Eq.~\eqref{quark_loop_factorised} can be estimated via $\delta L_{\Gamma, \Lambda_0}= L_\Gamma - L_{\Gamma, \Lambda_0}$ .
However, this additional cost can be significantly reduced by noticing that the inversions in the non-overlapping regions $\Lambda_0$ and $\Lambda_1$ can be done simultaneously
by generating multi-site Dirac sources $\chi$.
The number of simultaneous inversion depends only on the block size, and this results in a speed-up between $3-5$ for the measurements carried out in the bulk $24a \le x_0, y_0 \leq 71a$ with $\Delta/a=7-13$.
As the thickness $\Delta$ of $\Lambda_0$ and $\Lambda_1$ increases, the number of simultaneous inversions we can perform decreases.

This is far smaller than the cost required to achieve the same error reduction ($\approx 50$) for the unbiased estimator $C_\Gamma^{(d)}$ with the standard method, which would require roughly
a factor of $\approx 2500$ more statistics, corresponding to $\mathcal{O}(10^6)$ gauge configurations, which is computationally prohibitive.
\section{Conclusions}
In this work, we have introduced a new subtraction method for constructing biased estimators of quark-line disconnected two-point functions. The method exploits the locality of quark propagation through a domain decomposition of the Dirac operator. In particular, we subtract the completely factorised contribution obtained from quark loops restricted to non-overlapping regions of the lattice.
In the analysis with standard distillation, we find that the method trades a small bias for a substantial reduction of the statistical fluctuations. The subtracted contribution is expected to give only a negligible contribution to the physical correlator in channels where the relevant states have only a small gluonic component, while dominating its variance. This provides a practical alternative to completely neglecting disconnected contributions when their direct computation is prohibitively expensive.

We have tested the method for flavour-singlet charmonium observables and find a strong reduction of the statistical errors. For the largest block size considered, $\Delta/a=13$, the statistical error is reduced by a factor of almost $50$, corresponding to a variance reduction of $\mathcal{O}(2500)$, while the total computational cost increases only by approximately a factor of two.

The performance of the method is, however, channel and operator dependent.
For $\Gamma=\gamma_5$, the use of Gaussian profiles provides an additional reduction of the statistical uncertainties, and the subtraction method remains effective. 
In contrast, in the $\Gamma=I$ channel, and particularly for the narrowest profile $\omega_0=0.05$, a large fraction of the signal is contained in the completely factorised contribution. Neglecting this term therefore induces a sizeable bias. 
The large factorised contribution may indicate a sizeable gluonic component in these states, whose identification, however, requires a dedicated variational analysis.

Beyond charmonium, the method can also be applied to light-quark flavour-singlet observables. When the state of interest is the lightest state in the corresponding channel, the large-distance behaviour is more straightforward to interpret, without the ambiguity encountered in charmonium where lighter states can continue to pull the effective mass downward.

Importantly, the fact that both the signal and the statistical noise are concentrated in the completely factorised contribution makes this term particularly suitable for multilevel sampling. By computing this dominant contribution with a multilevel algorithm and the remaining, less noisy terms with standard sampling, one can reduce the statistical uncertainties while retaining an unbiased estimator of the full correlator. This approach avoids the need for the more elaborate approximation of the quark propagator involving additional terms employed in Ref.~\cite{Barca:2025dca}. It therefore provides a promising pathway towards efficient, unbiased estimators of quark-line disconnected correlators, including in channels where the subtraction approximation alone is not sufficient.

\section{Acknowledgments}
We thank the members of the FOR5269 research unit for useful discussions. 
The authors gratefully acknowledge the Gauss Centre for Supercomputing e.V. (www.gauss-centre.eu) for funding this project by providing computing time on the GCS Supercomputer SuperMUC-NG at Leibniz Supercomputing Centre (www.lrz.de) and the scientific support and HPC resources provided by the Erlangen National High Performance Computing Center (NHR@\allowbreak FAU) of the Friedrich-Alexander-Universität Erlangen- \allowbreak Nürnberg (FAU) under the NHR
project k103bf. NHR funding is provided by federal and Bavarian state authorities. (NHR@\allowbreak FAU) hardware is partially funded by the German Research Foundation (DFG) – 440719683. The work is supported by the German Research Foundation (DFG) research unit FOR5269 "Future methods for studying confined gluons in QCD". This project received funding from the European Research Council (ERC) via the project ”LEEX” grant agreement 101170304, funded by the European Union. Views and opinions expressed are however those of the author(s) only and do not necessarily reflect those of the European Union or the European Research Council Executive Agency (ERCEA). Neither the European Union nor the ERCEA can be held responsible for them.
J.A.U.-N. acknowledges support from a Research Ireland (Science Foundation Ireland) Frontiers for the Future Project award with grant number SFI-21/FFP-P/10186.

\bibliographystyle{unsrt}
\bibliography{bibliography}

\end{document}